\documentclass[conference]{IEEEtran}
\usepackage{url}
\usepackage{graphicx}
\usepackage{bbding}
\usepackage{pifont}
\usepackage{wasysym}
\usepackage{amssymb}
\begin{document}

\title{Setting the Foundation for Scientific Inquiry and Computational Thinking in Early Childhood using Lego Machines and Mechanism Education Kit }
\author{\IEEEauthorblockN{Vishnu Agrawal}
\IEEEauthorblockA{Robo-G, India\\
vishnu.agrawal19@gmail.com}
\and
\IEEEauthorblockN{Ashish Sureka}
\IEEEauthorblockA{Ashoka University, India\\
ashish.sureka@ashoka.edu.in}
}   
\maketitle

\begin{abstract}
We present a case study and an experience report on teaching engineering skills to young learners in the $7$ to $10$ years age group. Teaching engineering skills through a constructivist approach involving hands-on activities by designing and building machines and mechanism through concrete objects helps in developing the problem solving abilities of the child. Such activities also helps in laying the foundation for computer programming courses and scientific inquiry at elementary level. We present a learning environment and a curriculum using Lego simple machines and mechanism kit. Our objective is not only to teach concepts of basic machines but also develop essential skills like team work and collaboration, communication and time management in students belonging to the early childhood age group. We conduct experiments on a batch of five students during a summer camp and used a mixed method approach to collect both qualitative and quantitative data about their learning and behavior. Our research and findings provides empirical evidences that it is possible to develop engineering skills at early childhood level in addition to developing language literacy and mathematical thinking. 
\end{abstract}

\section{Research Motivation and Aim}
The curricula for early childhood education when kids are normally aged between $7$ to $10$ years and in Grade $1$ to $4$ has primarily focused on developing mathematical thinking and language literacy along with extra-curricular activities required for a holistic development in child. Engineering skills are not taught and setting the foundations for computer programming is not done in early childhood education. There have been some successful attempts made by educators to teach engineering skills, programming and computational thinking to elementary level children between the age group of $10$ to $14$ \cite{rogers2004bringing}\cite{vidushi2016}\cite{chaudhary2016experimental}\cite{bers2014computational}\cite{ruiz2004robotics}. For example, Tufts University Center for Engineering Educational Outreach (CEEO) has worked towards integrating engineering into elementary as well as high school education in schools belonging to the state of Massachusetts \cite{ rogers2004bringing}. Another example is of University of Chile at Santiago which has made substantial contributions towards teaching programming and engineering through robotics in various schools in Chile \cite{ruiz2004robotics}. There are several such examples all over the world.  

While there has been studies focusing on integrating engineering skills and teaching programming to children above the age of $10$, there are lack of research studies and experience reports from educators on the possibility of developing engineering and computational thinking in the early childhood age group (between $7$ to $10$ years and in Grade $1$ to $4$). Chambers et al. conduct a study with children of $8$ to $9$ years of age and teach them the concept of gears, power, speed and mechanical advantages \cite{chambers2008developing}. Their investigation demonstrates that such concepts can be taught to young learners through an exploratory and a hands-on methodology \cite{chambers2008developing}. We believe that many such case-studies are needed to understand the area well and our work presented in this paper confirms the findings of previous studies showing encouraging results. Children between the age of $7$ to $10$ are young learners and we hypothesize that engineering skills and computer programming can be taught to them. We believe that engineering skills can be integrated in their curriculum in such a way that they learn programming in a fun and playful manner. The motivation of the work presented in this paper is to investigate if we can develop engineering skills in children between the age of $7$ to $10$ and develop computational thinking skills in them so that the foundation for computer programming and scientific inquiry is laid. 

Lego\footnote{\url{https://education.lego.com/en-us}} has several education kits for various age groups and one of their kits is on simple machine and mechanism. Lego simple machine and mechanism kit contains more than $200$ bricks and elements that can be joined to create structures and mechanisms. The components in the set and support material can be used to teach the principles and functioning of various types of simple machines such as wheels and axles, pulleys, levers and gears. In addition to teaching the basic concepts of simple machines and exposing the students to engineering, our objective is to also examine if we can cultivate team work and collaboration, stress and pressure management, communication and time management skills in them. The specific aim of the work presented in this paper is to share our experiences on the application of Lego simple machines and mechanism education kit to teach basic engineering skills as well as other essential skills like team work and communication to young learners in the age of $7$ - $10$. 
\section{Research Contributions}
In context to existing research studies on early childhood education, the work presented in this papers makes the following novel contributions and fresh perspectives:
\begin{enumerate}
\item We define a curriculum to teach the concepts of simple machines and mechanism for children in Grade $1$ to $3$. We define a curriculum consisting of lessons, work sheets, projects, home works in which students learn the concepts of levers, wheels and axle, pulley, inclined plane, wedges, screws, gears, cam and ratchet.
\item We define a teaching methodology and structure the course so that the students acquire skills in team work and collaboration, oral and written communication, time management, self-discipline and self-management, creativity and imagination, problem solving, leadership and handing work related pressure and stress. The objective is to set the foundation for engineering and technology courses which requires teaching both technical and non-technical skills.
\item We conduct a case-study consisting of controlled experiments, making observations and collecting data about student behavior and learning. We share our experience as educators and mention the challenges we encountered and how we encountered the difficulties. Our experience report can serve as a recommendation and reference to early childhood educators interested in teaching engineering design and computational thinking to young learners and elementary level school students. To the best of our knowledge, the work presented in this paper is among the very few case-studies on using Lego simple machines and mechanism kit for early childhood education.
\end{enumerate}
\section{Proposed Curriculum and Pedagogy}
\begin{table*}[ht]
\begin{center}
\caption{Mapping of Topic and Skill. TWC: Team Work and Collaboration, OWC: Oral and Written Communication, TMM: Time Management, CRI: Creativity and Imagination, PRS: Problem Solving, LDS: Leadership Skills, SPM: Stress and Pressure Management, WS: Worksheets in Class, HW: Homework}
\begin{tabular}{ |c|c|c|c|c|c|c|c|c|c|c|c|c|c|c|c|c| }
\hline
& \multicolumn{2}{|c|}{\textbf{Lever}} & \multicolumn{2}{|c|}{\textbf{Wheels/Axle}} & \multicolumn{2}{|c|}{\textbf{Pulley}} & \multicolumn{2}{|c|}{\textbf{Planes}} & \multicolumn{2}{|c|}{\textbf{Wedges}} & \multicolumn{2}{|c|}{\textbf{Screws}} & \multicolumn{2}{|c|}{\textbf{Gears}} & \multicolumn{2}{|c|}{\textbf{Ratchets}} \\ \hline 
\textbf{Skills} & \textbf{WS} & \textbf{HW} & \textbf{WS} & \textbf{HW} & \textbf{WS} & \textbf{HW} & \textbf{WS} & \textbf{TH} & \textbf{WS} & \textbf{HW} & \textbf{WS} & \textbf{HW} & \textbf{WS} & \textbf{HW} & \textbf{WS} & \textbf{TH} \\ \hline
\textbf{TWC} & \checkmark & & \checkmark & \checkmark & \checkmark & & & \checkmark & \checkmark & \checkmark & & \checkmark & \checkmark & \checkmark & \checkmark & \checkmark \\ \hline 
\textbf{OWC} & \checkmark & & \checkmark & \checkmark & \checkmark & \checkmark & & \checkmark & & \checkmark & & \checkmark & \checkmark & \checkmark & \checkmark & \\ \hline 
\textbf{TMM} & \checkmark & & \checkmark & \checkmark & \checkmark & & & \checkmark & & \checkmark & & \checkmark & \checkmark & \checkmark & & \\ \hline 
\textbf{CRI} & \checkmark & & \checkmark & \checkmark & \checkmark & \checkmark & \checkmark & \checkmark & \checkmark & & \checkmark & \checkmark & & \checkmark & & \checkmark \\ \hline 
\textbf{PRS} & & \checkmark & & \checkmark & & & \checkmark & & & & \checkmark & & \checkmark & & \checkmark & \\ \hline 
\textbf{LDS} & \checkmark & & & \checkmark & \checkmark & \checkmark & & \checkmark & \checkmark & \checkmark & & \checkmark & \checkmark & & & \\ \hline 
\textbf{SPM} & \checkmark & \checkmark & & & \checkmark & & \checkmark & & \checkmark & & \checkmark & \checkmark & & & & \checkmark \\ \hline 
\end{tabular}
\end{center}
\label{skills1}
\end{table*}
\subsection{Lessons and Topics}
We first provided an introduction to the term simple machines to students and showed them few examples demonstrating how they help humans and solve their problems. We taught levers to students and demonstrated how and why levers help in lifting load by applying less force. The basic principle and concepts of rotation point, distance, force and load were explained. Wheel and axle systems were explained. The concept of circular motion was explained and examples of wheels and axles in everyday objects were shown. We talked about pulleys and explained the basic principles which makes lifting loads easier. The concept of force and distance using different sized pulleys were explained. We taught them about inclined planes and slanted surfaces. We taught the basic principles of wedges and screws. Students were also introduced to gears and showed how different toothed wheel can be combined to solve a task. Students were also familiarized with the concept of cam and ratchets. Overall the lessons reinforced the concept of simple machines through various types of mechanisms as force multiplying devices and how they can help us do work easily by applying less effort.  
\section{Exercises, Work Sheets, Home Works and Tests}
The learning environment we create is primarily based on the constructivist theory of teaching and learning. Constructivist philosophy consists of aspects like learning by doing, building structures and artefacts using concrete objects and materials and learning by actively designing and constructing \cite{alimisis2009constructionism}\cite{bers2002teachers}\cite{williams2007acquisition}. Marina et al. has applied a constructionist approach for teaching robotics for early childhood education \cite{bers2002teachers}. Williams et al. conduct a qualitative study on the application of Constructivist practice and approach in a robotics summer camp for middle school students \cite{williams2007acquisition}.

Table I displays the mapping between topics covered in class and the skill. We gave several homework (HW) and work-sheets or tests (WS) in class to students. The class and homework assignments were prepared keeping in mind the learning objectives. For example, we ensured that several homework and class activities develops team skills in students.
\section{Design of Experiments and Methodology}
\begin{table}[ht]
\begin{center}
\caption{Student Demographic Details. PEL: Prior Experience with Lego Education Kits}
\begin{tabular}{ |c|c|c|c|c| }
\hline
\textbf{Student} & \textbf{Age} & \textbf{Gender} & \textbf{Grade} & \textbf{PEL} \\
\hline
S1 & 9 & Male & 2 & No \\ \hline
S2 & 9 & Female & 2 & No \\ \hline
S3 & 8 & Male & 1 & No \\ \hline
S4 & 10 & Male & 3 & No \\ \hline
S5 & 8 & Male & 1 & No \\ \hline
\end{tabular}
\end{center}
\label{students}
\end{table}
Table II shows the demographic data of the five students in our class. The first author of the paper taught five students in a small classroom setting so that we can provide personal attention and conduct a research study which requires careful observation and data collection. Table II displays information about student's age, gender, grade and any prior experience with robotics education kits. Our class consisted of students from age $8$ to $10$ belonging to grade $1$ to $3$. Four out of the five students were male and only one student was female. None of the students had any prior experience with robotics education kit on simple machines and mechanism as well as on robotics programming. 
We use mixed method approach in our research methodology. We collect and analyze both quantitative and qualitative data. The quantitative data that we collect is the performance of students in the form of grades in work sheets, home works and tests which reflects their learning outcome and mastery of the concept. We gather qualitative data by carefully observing the behavior of the students in the class. We believe that using a mixed method approach in our case results in a more complete and comprehensive understanding of the phenomenon we are studying. 
\section{Observations and Results}
\begin{table}[t]
\begin{center}
\caption{Student Grades for Various Skills}
\begin{tabular}{ |c|c|c|c|c|c| }
\hline
\textbf{Skills} & \textbf{S1} & \textbf{S2} & \textbf{S3} & \textbf{S4} & \textbf{S5} \\
\hline
\textbf{TWC} & B & A & A & A & B \\ \hline
\textbf{OWC} & A & A & B & A & B \\ \hline
\textbf{TMM} & B & B & B & B & B \\ \hline
\textbf{CRI} & A & B & A & B & B \\ \hline
\textbf{PRS} & B & B & B & B & B \\ \hline
\textbf{LDS} & C & B & B & B & B \\ \hline
\textbf{SPM} & B & B & B & B & C \\ \hline
\end{tabular}
\end{center}
\label{grades}
\end{table}
Table III displays the consolidated grades of all the five students for the seven non-technical skills or soft skills on which they were evaluated. The grade is a weighted average of several tests, work sheets and home-work assignments. We did the grading on a three point scale represented by A (Good), B (Satisfactory or Meets Expectations) and C (Fair or Needs Improvement). Table III reveals that students were able to work in teams and collaborate. The students also showed good oral and written communication skills. We were able to cultivate creativity and imagination in students as they were able to create interesting designs and try different things on their own after providing some directions. We found that time management, self-organization, leadership skills and managing stress and pressure in a competitive situation requires more maturity and training.

We conducted a student survey and collected their feedback about the course. We observed that overall all the students were satisfied with the course and were enthusiastic. Students enjoyed and liked working in teams and expressed their interest in joint work. Students also liked the idea of rotating team leadership between them as it gives an opportunity to improve their speaking skills and communication. Students were thrilled about building and constructing objects. Based on our interaction with the students, we could clearly observe an improvement in their motor skills, creative thinking, and peer-to-peer communication. We could also see an improvement in social and emotional skills as the classroom and homework activities required working together in groups and teams. 
\section{Conclusion}
Our research provides evidences that it is possible to teach engineering skills to early childhood age group students and also cultivate several soft skills in them through constructivist approach and using Lego simple machines and mechanism kit. However, our dataset consisted of only five students and similar experiments needs to be conducted on more students to further strengthen the conclusions. We believe that more research is needed to investigate the impact of our training on the preparedness and readiness of students to learn computer programming as a follow-up course. All the students in our case-study were from middle to high income group, were from good performing schools and none of them were from families of low income group or disadvantaged children. Our research shows short-term improvements in various targeted technical and non-technical skills of the student but more investigation is required on the long-term impact of such summer camps on student’s academic achievements and personal growth. 
\bibliographystyle{IEEEtran} 
\bibliography{ic3} 
\end{document}